# The effect of triplet production on pair–Compton cascades in thermal radiation*

A. Mastichiadis,[1][†] R.J. Protheroe[2] and A.P. Szabo[2][‡]
[1] *Laboratory for High Energy Astrophysics, NASA/Goddard Space Flight Center, Greenbelt, MD 20771, USA*
[2] *Department of Physics and Mathematical Physics, University of Adelaide, Adelaide, South Australia 5005, Australia*

**ABSTRACT**
We calculate the spectrum of photons resulting from electromagnetic cascades through thermal radiation, and examine the consequences of including triplet production in these cascades. We assume that the cascade is one-dimensional, and we find that this approximation is justified in the present work for thermal radiation with temperature less than $10^{-3} mc^2$. Results are obtained for both monoenergetic and power-law primary spectra, and for a variety of path lengths. We find that triplet production is particularly important in electron–photon cascades through thermal radiation when the primary energy exceeds $10^5 m^2 c^4 / kT$ for propagation over small path lengths. The importance of triplet production decreases as the path length increases, and it has no effect on saturated cascades.

**Key words:** radiation mechanisms: miscellaneous – gamma rays: theory.

## 1 INTRODUCTION

Electron–photon cascading through radiation fields involving photon–photon pair production (PP) and inverse Compton scattering (ICS) governs the spectrum of high-energy radiation from a variety of astrophysical sources (Aharonian, Kirillov-Ugryumov & Vardanian 1985; Zdziarski 1986; Svensson 1987). Such cascades have already been applied in determining the spectra of $\gamma$-rays from energetic objects such as compact X-ray sources exhibiting very high-energy activity (Mastichiadis 1986; Protheroe & Stanev 1987) and active galactic nuclei (Mastichiadis & Protheroe 1990). They also occur in spatially varying anisotropic radiation fields, modifying the primary $\gamma$-ray spectra emitted from compact lueminous objects (Protheroe, Mastichiadis & Dermer 1992). In addition, such cascades may enable $\gamma$-rays above $10^{15}$ eV to be observed from extragalactic objects at distances far in excess of the pair production interaction length in the microwave background (Protheroe 1986). At even higher energies, cascades initiated as a result of pion photoproduction interactions of cosmic ray protons in the microwave background may result in an observable $\gamma$-ray background at TeV energies (Halzen et al. 1990).

However, at the highest energies inverse Compton scattering becomes less important than the process ($\gamma e \rightarrow 3e$) referred to here as triplet production (TP). The importance of triplet production in astrophysics has been emphasized by Mastichiadis, Marscher & Brecher (1986), Mastichiadis (1991) and Dermer & Schlickeiser (1991). As was shown in those papers, triplet production cannot be ignored in collisions where the photon energy in the electron rest frame exceeds $10^2 mc^2$, and triplet energy losses dominate over inverse Compton energy losses for energies greater than $10^6 mc^2$, where $m$ is the electron rest mass. Despite this, TP has not yet been incorporated into full cascade calculations.

In this paper, we consider electron–photon cascades in which TP acts both as a competitive energy loss mechanism to ICS and as a particle re-injection mechanism. In previous work (Mastichiadis, Szabo & Protheroe 1991) we considered the effects of TP on cascades through 'thin' slabs of thermal radiation of temperature $T$. We found that for $\Theta E > 10^2$, where $\Theta = kT$ and $E$ is the primary electron energy, both in units of $mc^2$, TP can have a significant effect on the spectrum of particles emerging from the slab. Here we extend this work to include 'thick' slabs as well as saturated cascades, i.e. where the energies of the photons are below the pair production threshold on the background radiation field. In Section 2 we briefly describe the physical processes that are important in this type of cascade. The method of calculation, which employs a combination of the Monte Carlo technique and the matrix doubling technique of Protheroe & Stanev (1993), is described in Section 3. The results are discussed in Section 4 and we give our conclusions in Section 5.

## 2 THE PROCESSES

For all three processes ($\alpha$ = ICS, TP, PP), the mean interaction length in an isotropic and uniform radiation field with temperature $\Theta$ is given by (assuming $E \gg 1$)

$$x_\alpha(E,\Theta)^{-1} = \int_{\varepsilon_\alpha}^\infty n(\varepsilon,\Theta) \int_{-1}^{\mu_\alpha} \frac{\sigma_\alpha(s)(1-\beta\mu)}{2} d\mu d\varepsilon, \qquad (1)$$

where $n(\varepsilon,\Theta)$ is the differential photon number density of the radiation field, $\sigma_\alpha(s)$ is the cross-section for process $\alpha$, $s$ is the square of the total centre of momentum frame energy in units of $m^2c^4$, $\mu = \cos\theta$ is the cosine of the interaction angle, and $\beta c$ is the velocity of the primary particle ($\beta = 1$ for PP). Here $E$ is the energy of the primary particle and $\varepsilon$ is the soft photon energy, both in units of $mc^2$. For ICS there is no threshold energy and hence $\varepsilon_{\rm ICS} = 0$ and $\mu_{\rm ICS} = 1$. The threshold conditions for PP and TP are $s > 4$ and $s > 9$ respectively, and the corresponding photon threshold energies are $\varepsilon_{\rm PP} = E^{-1}$ and $\varepsilon_{\rm TP} = 2E^{-1}$. Similarly, $\mu_{\rm PP} = 1 - 2(\varepsilon E)^{-1}$ and $\mu_{\rm TP} = 1 - 4(\varepsilon E)^{-1}$. We compare the mean interaction lengths as a function of $\Theta E$ for the three processes in Fig. 1(a). In each case, the mean interaction lengths have been scaled to the Thomson mean interaction length $x_{\rm T}(\Theta) = 1/[N(\Theta)\sigma_{\rm T}]$, where $N(\Theta) = \int n(\varepsilon,\Theta)d\varepsilon$ is the total number density of the radiation field, and $\sigma_{\rm T}$ is the Thomson cross-section. From Fig. 1(a) we can see that TP has a shorter mean interaction length than ICS for $\Theta E > 10^2$.

The mean energy loss rates for ICS and TP are given by

$$-\left\langle \frac{dE}{dt} \right\rangle = \frac{E}{t_\alpha(E,\Theta)} = \frac{\kappa_\alpha(E,\Theta)Ec}{x_\alpha(E,\Theta)}, \qquad (2)$$

where $\kappa_\alpha(E,\Theta)$ is the mean inelasticity for process $\alpha$ as a function of the primary electron's energy $E$ and the temperature of the radiation field $\Theta$. We have calculated $\kappa_\alpha(E,\Theta)$ for both ICS and TP using a Monte Carlo calculation, which will be described in Section 3.1. In the case of TP where there are three electrons (throughout this paper 'electrons' is taken to mean both electrons and positrons) in the final state, we have taken the total energy of the two lowest energy electrons to be the energy lost. The resulting energy loss time-scales, $t_\alpha(E,\Theta)$, are shown in Fig. 1(b) in units of the time-scale for Thomson scattering. For comparison, we have also added the interaction rate for PP to Fig. 1(b). We can see that, while TP interactions occur more frequently than ICS interactions for $\Theta E > 10^2$, TP energy losses only dominate over ICS energy losses for $\Theta E > 10^6$. This is because the mean inelasticity for TP is relatively small (Mastichiadis 1991) and falls off as $E^{-0.65}$ for $\Theta E > 10^2$, whereas for ICS the mean inelasticity is approximately constant over the same energy range.

## 3 METHOD OF CALCULATION

The application of Monte Carlo techniques to cascades dominated by the physical processes described above presents some difficulties, which we will try to address in the following sections. In Section 2 we saw that, while TP occurs more frequently than ICS for $\Theta E > 10^2$, relatively small amounts of energy are lost per interaction. Hence high-energy electrons will undergo many triplet interactions, in which very little energy is lost, until they undergo a catastrophic ICS. If one follows all the produced pairs, the computing time required is increased considerably.

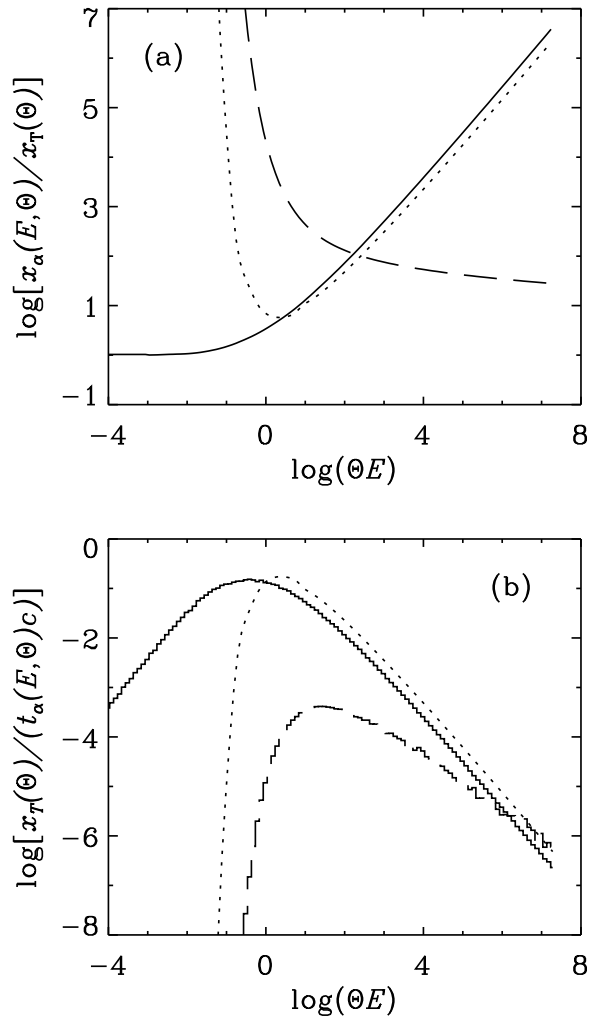

**Figure 1.** (a) The logarithm of the mean interaction length [scaled by $x_{\rm T}(\Theta)$] for ICS (solid curve), PP (dotted curve) and TP (dashed curve), as a function of $\log(\Theta E)$. (b) The reciprocal energy loss time-scale for ICS and TP [scaled by $c/x_{\rm T}(\Theta)$], as a function of $\Theta E$. Also shown is the interaction rate for PP. Curves are defined as in part (a).

The approach we use here is based on the matrix multiplication method described by Protheroe (1986) and subsequently developed by Protheroe & Stanev (1993). We use a Monte Carlo program to calculate the yields of secondary particles due to interactions with the thermal radiation. The yields are then used to build up transfer matrices which describe the spectra of particles produced after propagating through a slab of radiation with thickness $\Delta x$. Manipulation of the transfer matrices as described below then enables one to calculate the spectra of particles resulting from propagation through slabs of arbitrarily large thickness.

### 3.1 Particle yields

We define the quantity $Y_{ij}^{\alpha\beta}$ as the probability of producing a particle of type $\beta$ = e (electron) or $\gamma$ (photon) in the bin centred on energy $E_j$, when a primary particle with energy in the bin centred on $E_i$ undergoes an interaction of type $\alpha$ with the radiation field. To calculate $Y_{ij}^{\alpha\beta}$ we use a Monte Carlo simulation. For ICS and PP we have used the computer code described by Protheroe (1986, 1990), updated to model interactions with a thermal photon distribution of arbitrary temperature. For TP, we have used the code described by Mastichiadis (1991) to model the interaction, and we have sampled the initial conditions of the interaction as described by Mastichiadis et al. (1991). While calculating the yields for ICS and TP, we also calculated the mean inelasticity for ICS and TP used in Section 2 by summing over the fractional energy lost by the primary electron during each interaction.

### 3.2 Transfer matrix calculation

The transfer matrix, $T_{ij}^{\mu\nu}(\Delta x)$, gives the number of secondaries of type $\nu$ = e or $\gamma$, with energy in the bin centred on $E_j$, emerging from a slab of thermal photons of thickness $\Delta x$ when a primary of type $\mu$ = e or $\gamma$ and energy in the bin centred on $E_i$ is injected into the slab. To calculate the transfer matrices we have used a modification of the semi–analytical technique described by Protheroe & Stanev (1993). If $\Delta x$ is much shorter than the shortest interaction length in the cascade, i.e. $\Delta x \ll x_T$, then

$$T_{ij}^{ee}(\Delta x) \simeq \delta_{ij}\left[1 - \frac{\Delta x}{x_e(E_i)}\right] + \frac{\Delta x}{x_{ICS}(E_i)} Y_{ij}^{ICS\ e}$$
$$+ \frac{\Delta x}{x_{TP}(E_i)} Y_{ij}^{TP\ e}, \quad (3)$$

$$T_{ij}^{e\gamma}(\Delta x) \simeq \frac{\Delta x}{x_{ICS}(E_i)} Y_{ij}^{ICS\ \gamma}, \quad (4)$$

$$T_{ij}^{\gamma e}(\Delta x) \simeq \frac{\Delta x}{x_{PP}(E_i)} Y_{ij}^{PP\ e}, \quad (5)$$

$$T_{ij}^{\gamma\gamma}(\Delta x) \simeq \delta_{ij}\left[1 - \frac{\Delta x}{x_{PP}(E_i)}\right], \quad (6)$$

where $1/x_e(E_i) = 1/x_{ICS}(E_i) + 1/x_{TP}(E_i)$ gives the total mean interaction length for e$\gamma$ interactions, and $x_{ICS}(E_i)$, $x_{PP}(E_i)$ and $x_{TP}(E_i)$ are given by equation (1).

### 3.3 Matrix doubling method

The method used to calculate the spectrum of particles emerging from a slab of arbitrary thickness is described by Protheroe & Stanev (1993), and is summarized below. Once the transfer matrices have been calculated for a slab of thickness $\Delta x$, the transfer matrix for a slab of thickness $2\Delta x$ is simply given by applying the transfer matrices twice, i.e.

$$T_{ij}^{ee}(2\Delta x) = \sum_{k=j}^{i}\left[T_{ik}^{ee}(\Delta x)T_{kj}^{ee}(\Delta x)\right.$$
$$\left. + T_{ik}^{e\gamma}(\Delta x)T_{kj}^{\gamma e}(\Delta x)\right], \quad (7)$$

$$T_{ij}^{e\gamma}(2\Delta x) = \sum_{k=j}^{i}\left[T_{ik}^{ee}(\Delta x)T_{kj}^{e\gamma}(\Delta x)\right.$$
$$\left. + T_{ik}^{e\gamma}(\Delta x)T_{kj}^{\gamma\gamma}(\Delta x)\right], \quad (8)$$

$$T_{ij}^{\gamma e}(2\Delta x) = \sum_{k=j}^{i}\left[T_{ik}^{\gamma e}(\Delta x)T_{kj}^{ee}(\Delta x)\right.$$
$$\left. + T_{ik}^{\gamma\gamma}(\Delta x)T_{kj}^{\gamma e}(\Delta x)\right], \quad (9)$$

$$T_{ij}^{\gamma\gamma}(2\Delta x) = \sum_{k=j}^{i}\left[T_{ik}^{\gamma e}(\Delta x)T_{kj}^{e\gamma}(\Delta x)\right.$$
$$\left. + T_{ik}^{\gamma\gamma}(\Delta x)T_{kj}^{\gamma\gamma}(\Delta x)\right]. \quad (10)$$

The new matrices may then be used to calculate the transfer matrices for a slab of thickness $4\Delta x$, and so on. A slab of thickness $2^n\Delta x$ only requires the application of this 'matrix doubling' $n$ times. The spectrum of electrons and photons that escape from the slab is then given by

$$F_j^e(x) = \sum_{i=j}^{\infty}\left[T_{ij}^{ee}(x)F_i^e(0) + T_{ij}^{\gamma e}(x)F_i^\gamma(0)\right], \quad (11)$$

$$F_j^\gamma(x) = \sum_{i=j}^{\infty}\left[T_{ij}^{e\gamma}(x)F_i^e(0) + T_{ij}^{\gamma\gamma}(x)F_i^\gamma(0)\right], \quad (12)$$

where $F_i^e(0)$ and $F_i^\gamma(0)$ are the input electron and photon spectra (number of electrons or photons per unit energy in the bin centred on $E_i$), and $x = 2^n\Delta x$. In this way, cascades through thick slabs can be modelled quickly and efficiently.

The matrix method and matrix doubling technique have been used for many years in radiative transfer problems (van de Hulst 1963; Hovenier 1971), and in such problems scattering of the radiation means that 'reflection' in each layer $\Delta x$ must also be properly taken into account. Reflection is neglected in the present work because the angular deflection of the particles produced in the three interactions considered is negligible, and even the cumulative deflection after repeated interactions in the whole cascade is usually small, as we will show below.

### 3.4 Angular deflection of cascade particles

In this section, we justify neglecting the angular deflection of particles in interactions, and show that it is unnecessary to include reflection terms in the matrix calculations. We do not give here a rigorous treatment of the angular distribution of particles emerging from an interaction, as this is quite complicated. Instead, we make order of magnitude estimates of the maximum possible deflections. In practice, the deflections will usually be less than we estimate below. Generally, we will proceed along the following lines. We estimate the scattering angle per interaction, $\Delta\psi$, and then estimate the typical number of interactions required to 'cool' the particle, $N$. An estimate of the cumulative deflection after cascading is then $\psi \simeq N^{\frac{1}{2}}\Delta\psi$. Note that the number of scatterings, $N$, will in general be much smaller than the slab thickness over which we are calculating the cascade divided by the mean interaction length, except for very small slab thicknesses for which deflection will in any case be unimportant.

We start with a consideration of electrons with energies below the TP threshold. These electrons will cool exclusively by ICS, and the interactions will be in the Thomson regime. For scattering of a soft photon of energy $\varepsilon$ by a relativistic electron of energy $E$, the typical energy in the electron rest frame before scattering is $\varepsilon' \simeq E\varepsilon$. Since scattering is in the Thomson regime, this also equals the typical energy after scattering. The scattering is isotropic (within a factor of 2) in the electron rest frame, and so the transverse component of the photon's momentum after scattering is of order $\varepsilon'$. This will also be the order of magnitude of the transverse component of the electron's momentum after scattering, in both the electron rest frame and the laboratory (LAB) frame. Since the electron's energy loss per interaction is small when in the Thomson regime, the angular deflection per interaction is of order $\Delta\psi \simeq \varepsilon'/E \simeq \varepsilon$. In the present case of blackbody radiation of temperature $\Theta$, we have $\varepsilon \simeq 2.7\Theta$, and hence $\Delta\psi \simeq 2.7\Theta$. The typical energy loss of the electron per interaction is $E^2\varepsilon$. The number of interactions to 'cool' the electron is then typically $N \simeq E/E^2\varepsilon \simeq 1/2.7\Theta E$. Thus the cumulative scattering angle is

$$\psi \simeq N^{\frac{1}{2}}\Delta\psi \simeq \frac{1.6\Theta}{\sqrt{\Theta E}}. \tag{13}$$

As expected, this increases as the energy decreases. At the lowest energy used in the present work, for which $\Theta E = 10^{-4}$, the cumulative scattering angle will be small (less than $10°$) provided that $\Theta < 10^{-3}$ (i.e. for temperatures less than $6 \times 10^6$ K). In reality, the limit will be somewhat higher because our 'typical' deflections are in fact overestimates.

For PP well above threshold (i.e. $s \gg 4$), the deflection angles are quite small as we will show below. By integrating over the differential cross-section for PP (Jauch & Rohrlich 1976), we find that the median production angles in the centre of momentum (CM) frame of the two electrons are $\Delta\psi^* \simeq 2/s^{\frac{1}{4}}$ and $(\pi - 2/s^{\frac{1}{4}})$ with respect to the initial direction of the energetic photon in the CM frame. The angle between the energetic photon's direction in the CM frame and its direction in the LAB frame is of order $\sqrt{s}/E$ and so can be neglected in comparison with the median production angles. The transverse component of a produced electron's momentum is $p_\perp \simeq (\sqrt{s}/2)\sin\Delta\psi^* \simeq s^{\frac{1}{4}}$. The longitudinal component of momentum in the LAB frame is obtained by a Lorentz transformation using $\gamma_{\rm CM} \simeq E/\sqrt{s}$. We obtain $p_\parallel \simeq E(1-1/\sqrt{s})$ (corresponding to the high-energy particle with $\Delta\psi^* \simeq 2/s^{\frac{1}{4}}$) and $p_\parallel \simeq E/\sqrt{s}$ (corresponding to the low-energy particle). We see that usually the high-energy particle carries off most of the energy. The LAB frame production angles are then $\Delta\psi \simeq s^{\frac{1}{4}}/E$ (high-energy particle) and $\Delta\psi \simeq s^{\frac{3}{4}}/E$ (low-energy particle).

For ICS in the extreme Klein–Nishina regime, by integrating over the differential cross-section (Jauch & Rohrlich 1976) we find the median angle in the electron rest frame between the photon's direction after scattering and its initial direction to be $\Delta\psi' \simeq \sqrt{2}/(\varepsilon E)^{\frac{1}{4}}$. The angle between the photon's direction in the electron rest frame before scattering and the boost direction is $\sim \pi - 1/E$, and this may be neglected in comparison with the median scattering angle. The median LAB-frame angle between the scattered photon's direction and the initial direction of the electron is found by Lorentz transformation to be $\Delta\psi \simeq \sqrt{2}(\varepsilon E)^{\frac{1}{4}}/E$. With the help of further Lorentz transformations one obtains the median LAB-frame energies after scattering of the electron, $E_1 \simeq E/\sqrt{\varepsilon E}$, and of the photon, $\varepsilon_1 \simeq E(1 - 1/\sqrt{\varepsilon E})$, and the median LAB-frame electron scattering angle $\Delta\psi \simeq \sqrt{2}(\varepsilon E)^{\frac{3}{4}}/E$. Since, for ICS, $s = 1 + 2E\varepsilon(1-\mu)$, we find that the median LAB-frame angles and energies of the photon and electron after scattering are within a factor of 2 of those of the high-energy and low-energy electrons in pair production interactions well above threshold.

For ICS in the extreme Klein–Nishina regime, it is found that the high-energy particle taking most of the energy is almost always the scattered photon, and so in the cascade well above the PP threshold the high-energy particle alternates between an electron and a photon. The high-energy particle has negligible deflection in comparison with the low-energy particle. In the cascade, particles will migrate from high to low-energy by a variety of routes – in some collisions a particle will come out as the low energy particle, while in other it will come out as the high-energy particle. First we consider the cooling of the 'high-energy particle', i.e. we follow only the highest energy particle coming out of each interaction. The energy lost per interaction is $\Delta E \simeq E/\sqrt{s}$, giving $N \simeq \sqrt{s}$ collisions to cool the particle, with a resulting cumulative deflection of $\psi \simeq \sqrt{s}/E$. If instead we follow the low-energy particle produced, as it is effectively cooled in one interaction we have $\psi \simeq s^{\frac{3}{4}}/E$. This latter deflection imposes the stronger constraint and we see that, if the small-angle criterion is satisfied at some energy, it will automatically be satisfied at higher energies. We will postpone a discussion of the implications of this result until we discuss PP near threshold.

In the case of PP (and also TP) the largest deflection occurs close to threshold when the particles are produced almost isotropically in the CM frame. We shall consider the 'worst case scenario' of the produced particles sharing equally the energy available, and being produced in the CM frame with their velocity vectors transverse to the primary particle's direction. In this case, the energy of an electron in the CM frame is $\sqrt{s}/n$, where $n$ is 2 (3 for TP). Neglecting the electron mass, which will again cause overestimation of scattering angles, the transverse momentum (in any frame) is also $\sqrt{s}/n$. The longitudinal component of the LAB-frame momentum is $\gamma_{\rm CM}\sqrt{s}/n$, giving a deflection angle of $\Delta\psi \simeq \sqrt{s}/E$. Interactions such as these occurring near threshold will result in particles produced below the threshold for further cascading-type interactions (PP, or ICS in the Klein–Nishina regime), and in any case the mean interaction lengths near threshold are extremely large and so such interactions are rare. Hence $N \simeq 1$ and we obtain $\psi \simeq \sqrt{s}/E$. At threshold $\sqrt{s} = 2$, and therefore for small-angle deflections ($\psi < 10°$) we require $E > 12$. For PP, however, $\sqrt{s} = \sqrt{2E\varepsilon(1-\mu)} \simeq 3\sqrt{\Theta E}$ with head-on collisions ($\mu \simeq -1$) more likely at threshold, and so we obtain the constraint $\Theta < 0.03$ on the temperature of the radiation in order that angular deflections be small. With this temperature constraint, the small-angle criterion is clearly also satisfied in the high-energy part of the cascade, as discussed above both for the case of the deflection of the 'high-energy particle' and for individual scatterings when a particle emerges as the low-energy particle.

Finally, we consider TP well above threshold. An up-per limit to the deflection per interaction is given by $\Delta\psi < \Delta E/E$, where $\Delta E$ is the average energy lost per interaction. The number of interactions required to cool an electron is typically $N < E/\Delta E$. Hence $\psi < \sqrt{\Delta E/E} = \sqrt{x_{\rm TP}/t_{\rm TP}c}$. Taking the interaction length and energy loss time-scale for TP from Fig. 1, we see that, at energies where the TP interactions occur more frequently than IC ($\Theta E > 10^2$), $\psi$ is small.

We conclude that angular deflection is definitely unimportant in the present work provided that the temperature of the radiation is $\Theta < 10^{-3}$, and is probably unimportant to somewhat higher temperatures. As the temperature approaches $\Theta \simeq 1$, however, pair production by photon–photon collisions of photons of the radiation field, as well as interactions in any associated plasma, becomes important, and our results will not be valid. The neglect of reflection terms due to scattering in our matrix calculation is justified when $\Theta < 10^{-3}$. However, for higher temperatures (but still with $\Theta \ll 1$), our results presented below in terms of the spectra of particles or radiation emerging 'from a slab of thickness $x$' should instead be interpreted to mean spectra of particles or radiation 'at a time $x/c$ after injection of the primary particle into the radiation field'.

## 4 RESULTS AND DISCUSSION

We have calculated the photon spectrum emerging from slabs of various thicknesses for both monoenergetic primaries and a power-law primary spectrum. The power-law spectrum has the form $F_0(\Theta E) = A(\Theta E)^{-\Gamma}$, where $E_{\min} \le E \le E_{\max}$ and A is fixed by normalizing the spectrum to contain one particle. Results are given below for both electron and photon primaries.

### 4.1 Monoenergetic injection

The spectra of electrons and photons that escape from a relatively thin slab of thickness $2^5 x_{\rm T}(\Theta)$ are shown in Fig. 2. In (a) primary electrons of energy $E = 10^5/\Theta$ are injected into the slab, and we can see that the inclusion of TP causes a significant difference in both the shape and the normalization of the resultant spectra. Both the photon and the electron spectra are enhanced by as much as four orders of magnitude at energies below $E \sim 10^3/\Theta$. Similarly in Fig. 2(b) spectra are shown for primary electrons of energy $E = 10^7/\Theta$. Here the effect of including TP is even more prominent, with the spectra being enhanced by as much as eight orders of magnitude at low energies.

In Fig. 3 we investigate the effect of increasing the slab thickness on the escaping photon spectrum. We show in Fig. 3(a) the photon spectra obtained for electron primaries with an energy of $E = 10^5/\Theta$ after cascading through slabs of thickness $2^{10}x_{\rm T}(\Theta)$ (thinnest curves), $2^{15}x_{\rm T}(\Theta)$ and $2^{20}x_{\rm T}(\Theta)$ (thickest curves). In each case the results with TP included (solid curves) are compared with those where TP is not included (dotted curves). From Fig. 3(a) we can see that, while the slab is 'thick' ($2^{10}x_{\rm T}$ and $2^{15}x_{\rm T}$ for this primary energy), the inclusion of TP can cause the normalization of the photon spectrum below the PP threshold to be enhanced by up to several orders of magnitude. However, once the cascade approaches saturation (e.g. $2^{20}x_{\rm T}$ for this primary energy) the results for the two cases become identical below the PP threshold. Similar behaviour can be seen in Fig. 3(b) where we show the photon spectra emerging from slabs of thickness $2^{15}x_{\rm T}(\Theta)$ (thinnest curves), $2^{20}x_{\rm T}(\Theta)$ and $2^{25}x_{\rm T}(\Theta)$ (thickest curves) for primary electrons of energy $E = 10^7/\Theta$. Here the inclusion of TP may enhance the photon spectrum at low energies by as much as four orders of magnitude. In each case, the cascade is seen to saturate for smaller slab thicknesses when TP is included, and there is also a significant difference in the shape of the spectrum between the PP threshold and the primary electron's energy.

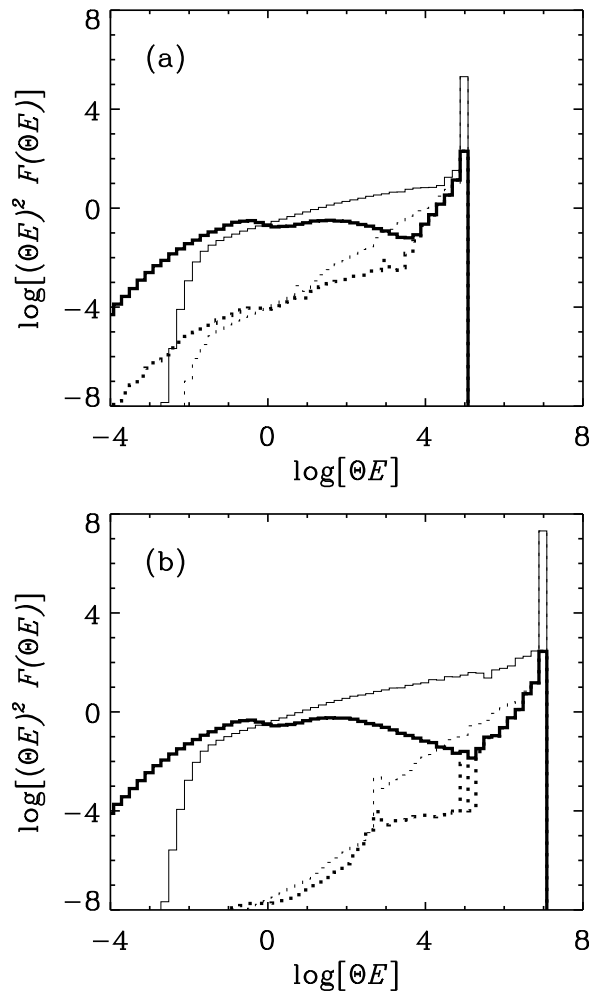

**Figure 2.** The spectrum of particles and radiation emerging from a slab of thickness $2^5 x_{\rm T}(\Theta)$ for monoenergetic primaries. Note that, because of the way in which $E$ and $\Theta$ are defined, $(\Theta E)^2 F(\Theta E)$ is dimensionless. Results are shown for the case where TP is included (solid curves) and the case where it is not (dotted curves). The thick (thin) curves represent the spectra of photons (electrons and positrons) that escape from the the slab. (a) Primary electron of energy $E = 10^5/\Theta$. (b) Primary electron of energy $E = 10^7/\Theta$.

We have also calculated the photon spectrum produced when photon primaries instead of electrons are injected into the slab. Results for slabs of thickness $2^{15}x_T(\Theta)$ (thinnest curves), $2^{20}x_T(\Theta)$ and $2^{25}x_T(\Theta)$ (thickest curves) and $E = 10^7/\Theta$ are shown in Fig. 3(c). Again TP is seen to have an important effect for cascades through 'thick' slabs. The magnitude of the enhancement below the PP threshold is less than for the equivalent case using electron primaries because the $\gamma$-rays pair-produce, and hence the mean energy of the first electrons and positrons produced is less than $E = 10^7/\Theta$.

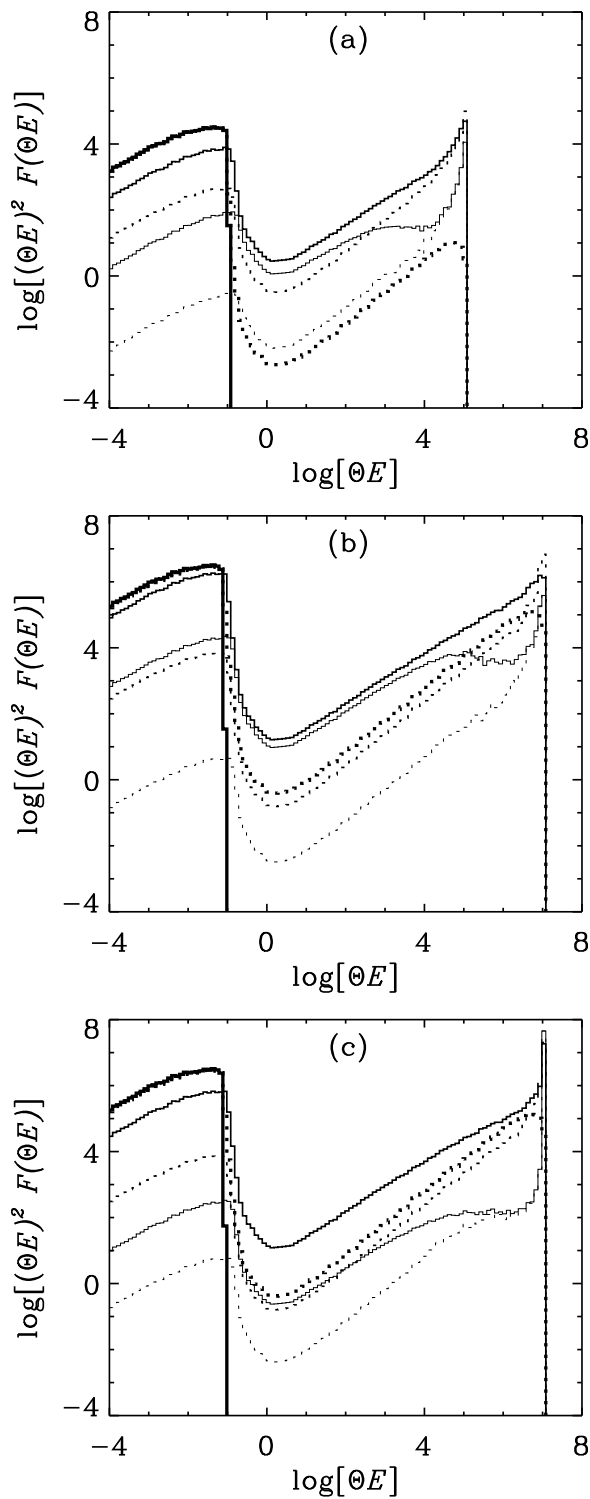

**Figure 3.** The spectrum of radiation emerging from slabs of various thicknesses for monoenergetic primaries. Results are shown for the case where TP is included (solid curves) and the case where it is not (dotted curves). (a) Primary electron of energy $E = 10^5/\Theta$ for slabs of thickness $2^{10}x_T(\Theta)$ (thinnest curves), $2^{15}x_T(\Theta)$ and $2^{20}x_T(\Theta)$ (thickest curves). (b) Primary electron of energy $E = 10^7/\Theta$ for slabs of thickness $2^{15}x_T(\Theta)$ (thinnest curves), $2^{20}x_T(\Theta)$ and $2^{25}x_T(\Theta)$ (thickest curves). (c) As for part (b), but with photon primaries.

The above results can be understood qualitatively as follows. Any electron in the system with $E\Theta \gg 10^2$ will interact with the ambient photons predominantly by TP interactions. As was emphasized by Mastichiadis (1991), in the absence of ordinary PP these interactions will tend to steepen the resulting photon spectrum and enhance the normalization over the 'pure' ICS case. The inclusion of PP should not change the above picture as long as the slab is not too thick. For this type of slab, $\gamma$-rays have a finite probability of escaping. However, those $\gamma$-rays that interact produce an electron–positron pair which will produce a number of secondary pairs through TP interactions. The secondaries are produced with rather low energies, and as such they will cool by ICS, producing additional photons below the PP threshold. As a result, the final photon spectrum will have a normalization that will be higher than that expected in the usual ICS–PP cascades. Furthermore, one would expect the ratio of the two normalizations to be given by the pair multiplicity for TP (the number of secondaries produced per primary electron) as defined by Mastichiadis (1991). Therefore, by increasing the primary energy, one would expect the difference from the usual cascades to be more significant. The Monte Carlo results given above seem to support this notion.

The situation changes when we consider saturated cascades. In this case, the inclusion of TP should not make any difference as the particular characteristics of the cascade are totally washed away. Both pair-producing processes, despite their different properties, will effectively degrade all high-energy particles and photons to energies below threshold. Therefore the end result should be the same irrespective of the way in which the particles lose their energy inside the slab.

It is evident that in going from moderately thin slabs, where the inclusion of TP can produce a dramatic difference of several orders of magnitude over the usual cascades, to very thick slabs, where the two cases are practically identical, one has a range of slab thicknesses where the inclusion of TP is of decreasing importance.

### 4.2 Power-law injection spectrum

The results for a power-law primary electron spectrum with $\Gamma = 1.5$, $E_{\min} = 10^{-4}/\Theta$ and $E_{\max} = 10^7/\Theta$ are shown in Fig. 4(a). Escaping photon spectra are shown for slabs of thickness $2^{15}x_T(\Theta)$ (thinnest curves), $2^{20}x_T(\Theta)$ and $2^{25}x_T(\Theta)$ (thickest curves). Again an enhancement in the low-energy photon spectrum is seen, although in this case the inclusion of TP only increases the normalization by a factor of $\sim 3$ for $2^{15}x_T$ and $2^{20}x_T$. As for the case of monoenergetic injection, once the cascade becomes saturated ($\sim 2^{25} x_T$ in this case) the low-energy photon spectra become identical. We have also calculated results for a steeper primary electron spectrum ($\Gamma = 2.0$), which are shown in Fig. 4(b). In this case, the inclusion of TP has little effect on the results, as most of the low-energy photon spectrum is dominated by cascades induced by primary electrons with energy below $E \sim 10^6/\Theta$ where ICS dominates the energy losses.

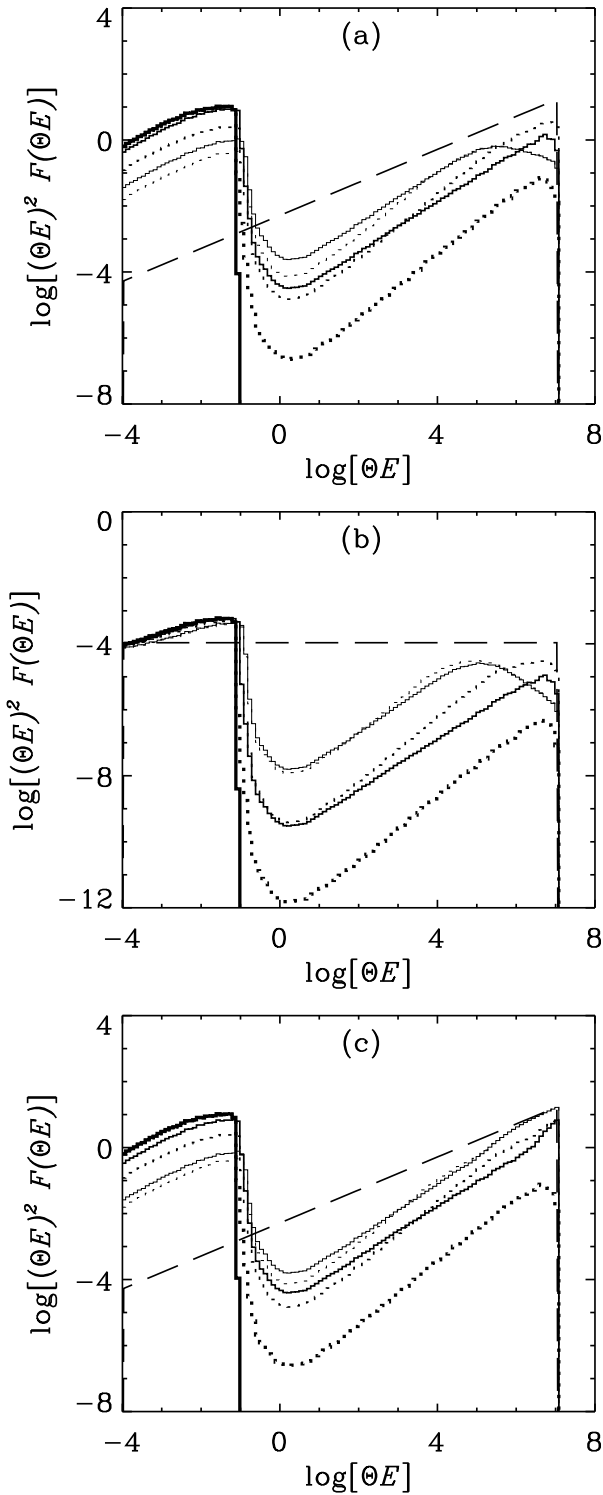

**Figure 4.** The spectra of radiation emerging from slabs of various thicknesses for a power-law primary spectrum of the form $F_0(\Theta E) = A(\Theta E)^{-\Gamma}$, $E_{\min} \leq E \leq E_{\max}$. Results are shown for the case where TP is included (solid curves) and the case where it is not (dotted curves). In each case the input spectrum is indicated by the dashed curve. (a) Primary electron spectrum with $\Gamma = 1.5$, $E_{\min} = 10^{-4}/\Theta$ and $E_{\max} = 10^7/\Theta$ for slabs of thickness $2^{15} x_T(\Theta)$ (thinnest curves), $2^{20} x_T(\Theta)$ and $2^{25} x_T(\Theta)$ (thickest curves). (b) As for part (a), except with $\Gamma = 2.0$. (c) As for part (a), but with photon primaries.

The escaping photon spectra that result when a power-law photon spectrum ($\Gamma = 1.5$, $E_{\min}$ and $E_{\max}$ as before) is injected into the slab are shown in Fig. 4(c). While an enhancement below the PP threshold is observed, its magnitude is again less than for the equivalent case with electron primaries.

## 5 CONCLUSIONS

In the present paper we have examined the effect of triplet production (TP) on electron–photon cascades through a thermal radiation field. These cascades have until now been examined by assuming that the only relevant processes are inverse Compton scattering (ICS) and photon–photon pair production (PP). By applying a Monte Carlo code and subsequently the matrix doubling technique, we have shown that the inclusion of TP can have significant effects on the emerging photon spectra. The importance of TP becomes more pronounced for high primary energies and for relatively thin slabs. The differences in the spectra occur both below and above the PP threshold. As the slab thickness increases the differences below the PP threshold diminish, and for the limiting case of an infinite slab (i.e. a saturated cascade) the inclusion of TP will have no practical consequences. However, we find that the inclusion of TP causes saturation to occur for smaller slab thicknesses, and so even for ultra-thick slabs, where the spectra are almost identical below the PP threshold, there will still be significant differences above the PP threshold. The significance of TP in astrophysical environments will thus depend solely on the individual source parameters, which in turn determine whether a saturated cascade occurs.